# Achieving room temperature plasticity in $TiO_2$ ceramics by flash processing

Xinghua Su [1] *, Zhenyu Li [1], Xufei Fang [2], Wenjun Lu [3], Mengyao Wang [1], Peng Zhao [1]

[1] School of Materials Science and Engineering, Chang'an University, Xi'an 710061, China

[2] Institute for Applied Materials, Karlsruhe Institute of Technology, Karlsruhe 76131, Germany

[3] Department of Mechanical and Energy Engineering, Southern University of Science and Technology, Shenzhen 518055, China

* Corresponding authors: suxinghua@chd.edu.cn (X. Su)

**Abstract:**

Ceramics are indispensable materials for modern engineering and high-technology applications. However, the intrinsic brittleness at room temperature can cause catastrophic failure and severely limit their widespread applications. Here, we report a flash processing (FP) method based on the coupling of electric and thermal fields to significantly improve the room-temperature plasticity of $TiO_2$ ceramics with a large plastic compressive strain of 15.4% during micropillar compression. This large room-temperature plasticity is attributed to the proliferation of stacking faults and twin boundaries during the loading process, which may be assisted by the high concentration of oxygen vacancies introduced by the FP method. Our findings suggest that FP can be a simple and effective method for achieving improved room-temperature plasticity of ceramic materials through defect engineering.

## 1. Introduction

Advanced ceramics are indispensable materials for modern engineering and high-technology applications, owing to their superior mechanical strength, chemical stability, and high-temperature capability [1-5]. However, the widespread structural application of ceramics is fundamentally bottlenecked by their inherent catastrophic brittleness at room temperature. Unlike metals, the strong ionic and covalent bonds in ceramics often impose a very high Peierls-Nabarro stress, which severely hinders dislocation mobility at room temperature, even if dislocation nucleation can be circumvented [6,7]. Consequently, under external mechanical loading, stress concentration at the pre-existing flaws, dislocation pileups, or crack tips inevitably triggers premature, catastrophic failure before occurrence of appreciable plastic deformation. Overcoming this intrinsic brittleness and achieving room-temperature plasticity has remained a long-standing challenge.

Over the past decades, intensive efforts have been dedicated to mitigating ceramic brittleness, primarily through extrinsic toughening mechanisms such as phase transformation [8], grain refinement [9], and the introduction of secondary phases [10]. While these strategies successfully enhance fracture toughness, they rarely alter the intrinsic deformation capability of the ceramic bulk itself. Intrinsic plastic deformation in ceramics typically requires high temperatures (usually above half of their melting points) to thermally activate slip systems, dislocation motion, and grain boundary sliding [11-13]. As a result, defects engineering, which involves introducing abundant defects (particularly dislocations) into ceramics, has been presented to achieve plasticity in ceramics at room temperature [14-16]. For example, Dong et al. achieved the exceptional tensile ductility exceeding 39.9% in $La_2O_3$ ceramics at room temperature using a concept of “borrowed-dislocations”, which entailed continuously transmitting dislocations from Mo metals and introducing them into $La_2O_3$ ceramics [14]. Fang et al. reported the room-temperature plasticity up to 32% in $SrTiO_3$ through mechanically seeded dislocations [15]. Shen at al. proposed an elevated temperature preloading method to introduce abundant defects into the $TiO_2$ ceramics, and achieved a 10% compressive strain at room temperature [16]. However, these approaches for introduction of defects require stringent pre-deformation or processing conditions, and are limited to specific material systems. Challenges in scalability remain. Therefore, it is necessary to develop more generalized methods to improve the plasticity of ceramics at room temperature.

Recently, the emergence of a novel field-assisted sintering technique, namely flash sintering, has revolutionized the synthesis and sintering of ceramic materials [17-20]. By applying a critical electric field accompanied by heating, flash sintering enables ultra-fast densification and solid-state reaction within seconds at significantly lowered furnace temperatures. For example, Cologna et al. demonstrated that yttria-stabilized zirconia (YSZ) could be rapidly densified to full density within only a few seconds at 850 °C via flash sintering [17]. More remarkably, Liu et al. reported that ZnO ceramics could be densified by flash sintering even at room temperature [21]. Recently, Su et al. successfully fabricated dense high-entropy rare-earth zirconate ceramics via flash sintering of a mixture of oxide powders at 865 °C for 60 s, during which solid-state reactions and densification proceeded simultaneously [22]. Beyond its unprecedented efficiency in sintering, the coupling of electric and thermal fields has been found to induce extraordinary non-equilibrium states, including massive point defect generation, accelerated diffusion kinetics, and altered grain boundary energy [23-28]. Emerging evidence suggests that these electro-thermal coupled fields can profoundly interact with the defect chemistry and defects kinetics of ceramics. However, the exploration of flash technique has been predominantly confined to the densification stage. Whether such a unique electro-thermal coupling can be exploited to fundamentally reshape the mechanical response, specifically to unlock room-temperature plasticity, remains largely unexplored.

In this work, we report an electro-thermal coupled flash processing (FP) strategy to significantly improve the room-temperature plasticity of ceramics, as showcased on $TiO_2$. Differs from the flash sintering, FP operates directly on already densified ceramics rather than porous green bodies. Moreover, instead of driving densification, the FP is leveraged for microstructural tuning via defect engineering, providing a versatile platform to tune the physical-chemical properties of ceramic materials. Titanium dioxide ($TiO_2$), a classic transition metal oxide with high industrial relevance, typically exhibits a highly brittle nature under ambient conditions. The brittle to ductile transition temperature of bulk $TiO_2$ crystals was reported to be higher than 600 °C [29], much higher than room temperature. Here in this work, we show that after FP, $TiO_2$ demonstrates substantial plastic strain under mechanical loading at room temperature, completely deviating from the conventional brittle fracture behavior.

Detailed structural characterization reveals that the $TiO_2$ after FP contains a minor fraction of pre-existing stacking faults and twins, alongside a high concentration of oxygen vacancies. During room-temperature compression, the multiplication of these stacking faults and twin boundaries drive the significant room-temperature plasticity. The proliferation of these planar defects is strongly assisted by the abundant oxygen vacancies, which likely facilitate their nucleation and growth. Different from the previously reported dislocation-assisted room-temperature plasticity, our findings uncover a novel oxygen-vacancy-assisted stacking fault and twinning proliferation route for improving ceramic plasticity. Our findings not only suggest that FP can be a facile and effective method to improve the room-temperature plasticity of ceramic materials, but also helpful to deeply understand the deformation mechanisms of ceramic materials.

## 2. Materials and methods

### 2.1 Flash processing

Rutile $TiO_2$ bulk samples (relative density of 98.2%) with a diameter of ~6.2 mm and a thickness of ~2.8 mm were used as pristine samples. The samples were first ground with 4000-grit silicon carbide abrasive paper and then polished with a diamond suspension. After that, ultrasonic cleaning was performed for 10 min to remove contaminants. The dense $TiO_2$ samples were sandwiched between two Cu sheets as electrodes. To ensure rigid contact between the sample and the electrode, a pressure of approximately 50 kPa was applied to the electrodes through the alumina plates and rods. A DC power supply (HPS0614, Hangzhou Lanyi Electronics Co., Ltd., China) was used to apply a constant electric field of 500 V $cm^{-1}$ to the samples. Then the samples were heated in a furnace at a rate of 10 °C $min^{-1}$ in air. When the current flowing through the sample increased rapidly to preset values of 230 and 280 mA (corresponding to a current density of 0.762 and 0.928 A $cm^{-2}$, respectively), which signaled the occurrence of the flash event. At the same time, the power supply was switched from the voltage-control mode to the current-control mode. The sample was then kept at a constant current state for 60 s. After that, the power supply and furnace were turned off, and the samples were furnace cooled to room temperature.

### 2.2 Micropillar compression test

$TiO_2$ micropillars with ~3 μm in diameter and ~6 μm in height were prepared from the $TiO_2$ sample near the anode side using a Thermo Fisher Scios 2 Dual Beam Scanning Electron Microscope-Focused Ion Beam (SEM-FIB). The micropillars were prepared under the following sequence of FIB parameters: rough cutting at 30 kV and 7 nA, fine polishing at 30 kV and 50 pA, and final low-voltage cleaning at 5 kV and 23 pA to minimize surface damage. In situ micropillar compression tests were performed using a nanomechanical testing system (FT-NMT04, FemtoTools) in displacement-controlled mode at a constant strain rate of $1\times10^{-4}$ $s^{-1}$.

**2.3 Sample characterizations**

Microstructure characterization was performed on a transmission electron microscope (TEM, Thermofisher Talos F200X) operated at 200 kV. TEM samples of $TiO_2$ with FP were prepared through the conventional cutting, grinding and ion milling. For samples post-micropillar compression, cross-section TEM samples were fabricated using a Thermo Fisher Scios 2 Dual Beam SEM-FIB system. The crystalline structure was identified by an X-ray diffractometer (XRD; D8 Davinci, Bruker, Germany) with Cu Kα radiation. Electron paramagnetic resonance (EPR) spectroscopy was carried out to detect oxygen vacancies using an A300 spectrometer (Bruker, Germany). The X-ray photoelectron spectroscopy (XPS) was performed on an ESCALAB 250 spectrometer (Thermo VG Scientific, USA) with monochromatic Al Kα excitation. It should be noted that the XRD, EPR and XPS were carried out on the region near the anode side of the sample.

**3. Results**

The schematic of flash processing (FP) is illustrated in Fig. 1. Initially, a constant electric field of 500 V $cm^{-1}$ was applied to the dense $TiO_2$ samples. Subsequently, the $TiO_2$ samples were heated in air. Upon reaching a temperature of ~575 °C, the electrical conductivity of the sample increased non-linearly, causing the current passing through the sample to surge rapidly to a preset limit (230 and 280 mA, corresponding to a current density of 0.762 and 0.928 A $cm^{-2}$, respectively). Thereafter, the sample was held under this constant current for 60 s (Fig. 1(b)). After the FP, the color of the $TiO_2$ sample became slightly darker, as shown in the inset of Fig. 1(b).

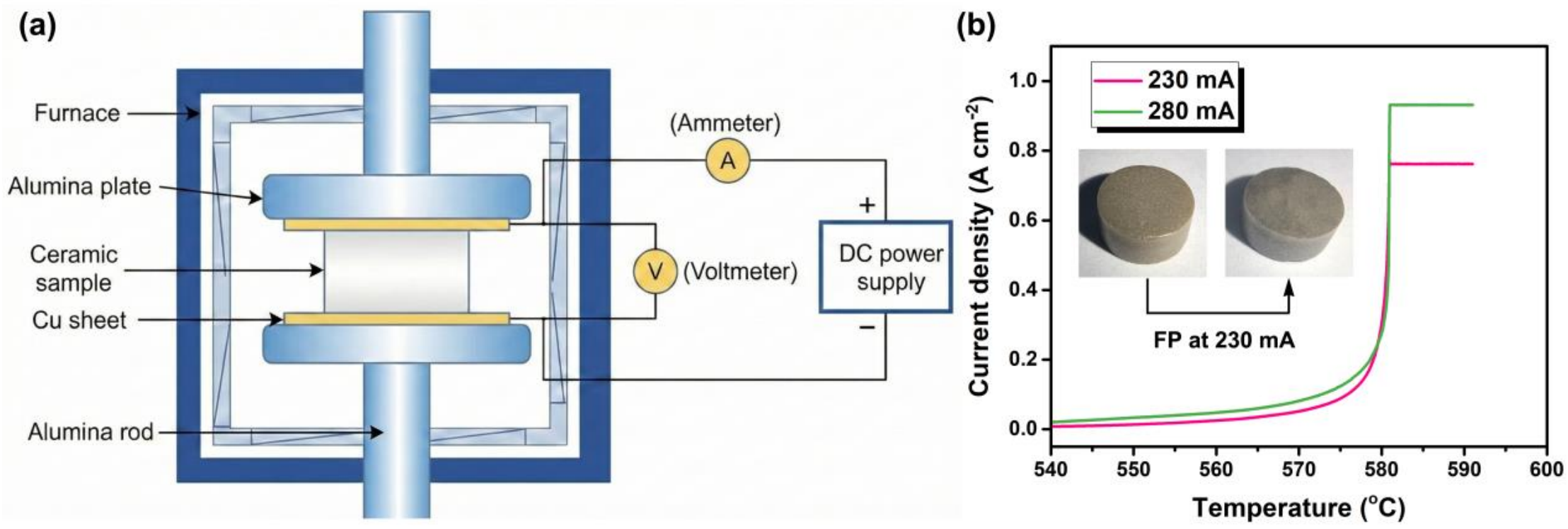

**Fig. 1** (a) Schematic experimental setup for FP. (b) Current density as a function of temperature during FP. The inset shows photographs of the $TiO_2$ samples before and after FP at 230 mA.

Micropillar compression tests of $TiO_2$ samples without FP and with FP at 230 and 280 mA were performed at room temperature with a constant strain rate of $1\times10^{-4}$ $s^{-1}$. For the pristine $TiO_2$ sample without FP, pillars experienced brittle catastrophic failure at an engineering strain of ~2.3% and an engineering stress of 3.6 GPa (Fig. 2(a-d)). The crack expansion was instantaneous and catastrophic (see supplementary Movie S1 for detail). In contrast, pillars underwent plastic deformation for the $TiO_2$ sample with FP at 230 mA (Fig. 2(e-h)). Serrations are clearly observed on the stress-strain curves, which are due to significant plastic deformation prior to failure. A maximum strain of ~15.4% (red curve in Fig. 2(e)) is achieved without noticeable major cracks. From the in situ SEM snapshots of a representative pillar, it can be seen that shear bands are formed and propagated throughout the entire pillar under successive compression (see supplementary Movie S2 for detail).

Interestingly, for the $TiO_2$ sample with FP at 280 mA, the stress-strain curves exhibit a pronounced work-hardening behavior (Fig. 2(i-l)), which is rarely observed in ceramic materials during deformation. The flow stress is increased continuously to 3.5 GPa under loading, and the engineering strain reaches 14.7% without brittle fracture (see supplementary Movie S3 for detail). Similarly, the generation and proliferation of shear bands occur during the compression process (Fig. 2(j-l)).

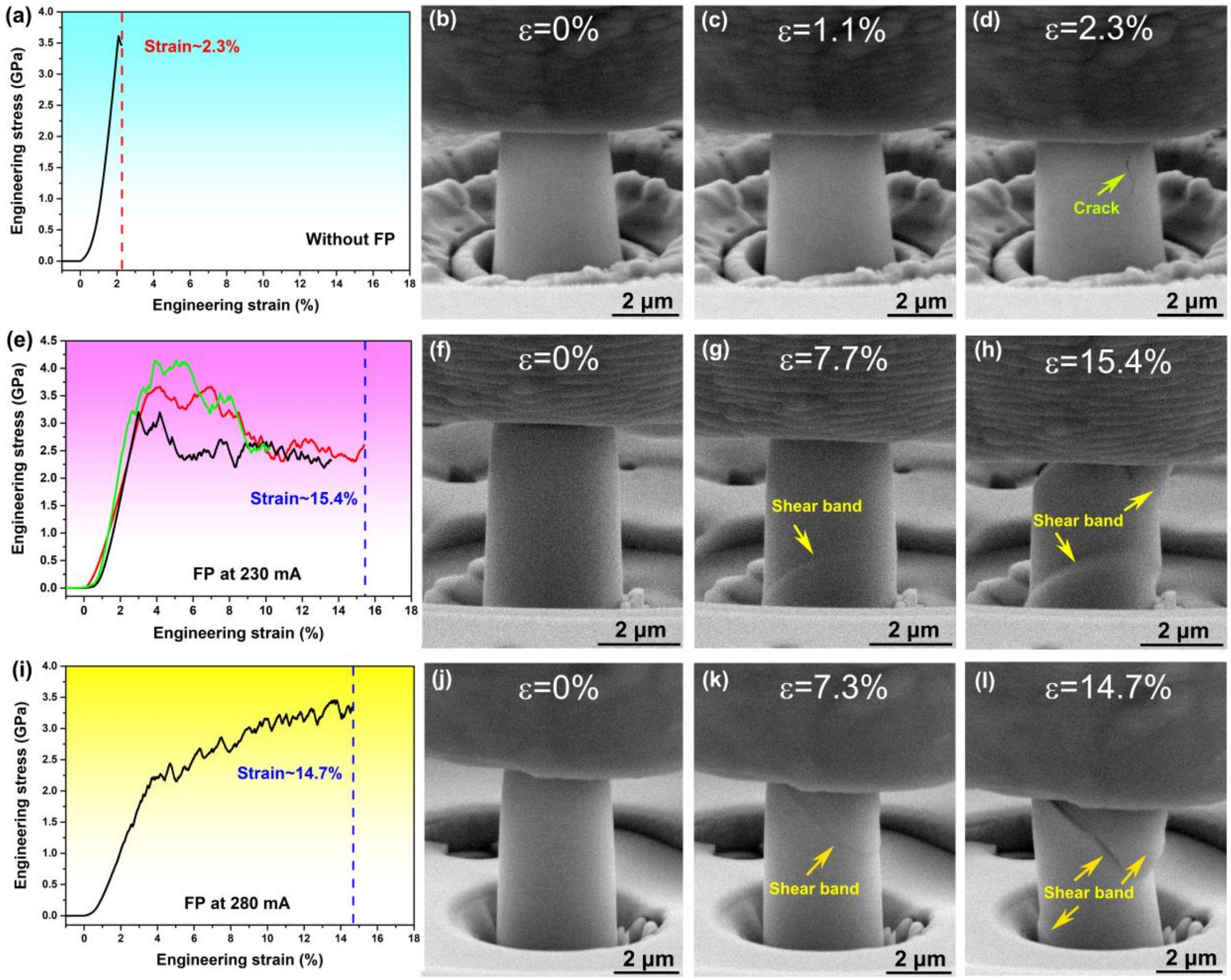

**Fig. 2** Uniaxial in situ microcompression tests at room temperature. (a-d) Stress-strain curve of $TiO_2$ without FP. (e-h) Stress-strain curves of $TiO_2$ with FP at 230 mA from different individual pillars. (i-l) Stress-strain curve of $TiO_2$ with FP at 280 mA.

To investigate the microstructural evolution during compressive deformation, TEM characterization was performed on $TiO_2$ samples (FP at 280 mA) subjected to compressive strains of 8.0% and 14.7%. For the pillar compressed to a strain of 8.0%, several parallel slip bands are observed (Fig. 3(a)). To further examine the microstructure in detail, the region highlighted by the red box in Fig. 3(a) is selected for TEM characterization. It is revealed that these parallel slip bands consist of stacking faults and twin boundaries, which is confirmed by the spot splitting and streaking lines in the selected-area electron diffraction (SAED) pattern (Fig. 3(b)). High-resolution TEM (HRTEM) image of the region in Fig. 3(b) (indicated by the red box) shows that stacking faults are localized at the twin boundary, which is further confirmed by the inserted fast Fourier transformation

(FFT) pattern (Fig. 3(c)).

For the pillar compressed to a strain of 14.7%, a high density of parallel slip bands is observed in the deformed region (Fig. 3(d)). In addition, no cracks are observed. These abundant slip bands intersected with the surface of the pillar and resulted in prominent shear offsets, as shown in Fig. 2(l). In Fig. 3(e), it reveals that these parallel slip bands are composed of an abundance of stacking faults and twin boundaries, as indicated by the inserted SAED pattern. The HRTEM image in Fig. 3(f) further confirms the co-existence of stacking faults and twin boundaries.

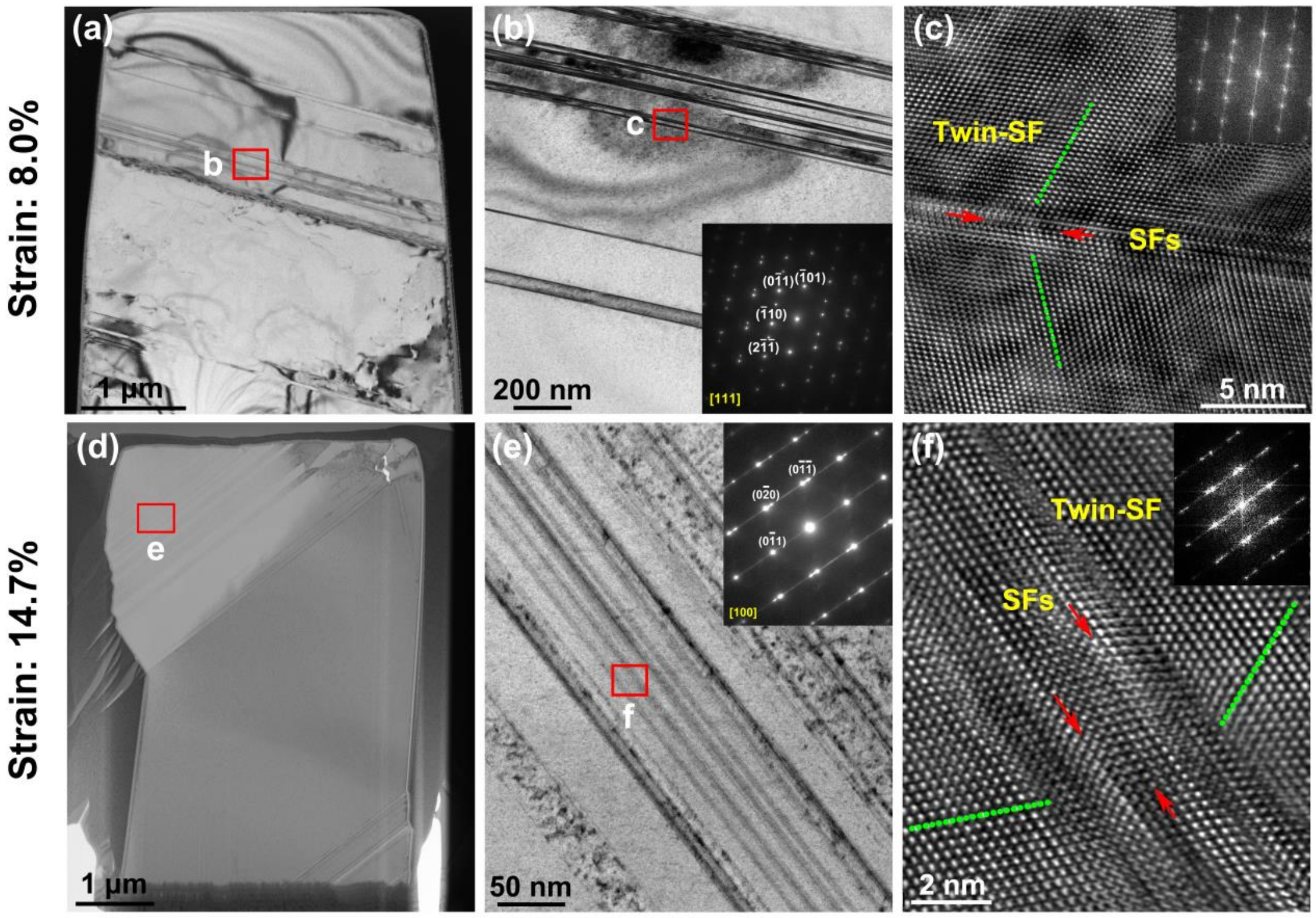

**Fig. 3** TEM images of $TiO_2$ (FP at 280 mA) pillars after compression tests at different strains. (a,b) Low-magnification TEM image of a pillar compressed at a strain of 8.0%. (c) HRTEM image. (d,e) Low-magnification TEM image of a pillar compressed at a strain of 14.7%. (f) HRTEM image.

To investigate the microstructural changes of $TiO_2$ before and after FP, TEM characterization was conducted on both samples. Figure 4 presents the TEM image of the sample with FP at 280 mA. Notably, the TEM sample was prepared via conventional approach: a slice was sectioned from the central part of the cylindrical $TiO_2$ sample, followed by mechanical grinding and ion milling. It can be clearly seen that the sample contains a small number of stacking faults (Fig. 4(a,b)). HRTEM image shows that twin boundaries are also formed, which are decorated with stacking faults (Fig. 4(c)). However, for the pristine $TiO_2$ sample without FP, TEM characterization reveals clean grain interiors, virtually free of defects (see supplementary Fig. S1). These observations indicate that a small number of stacking faults and twins were generated in the sample after FP.

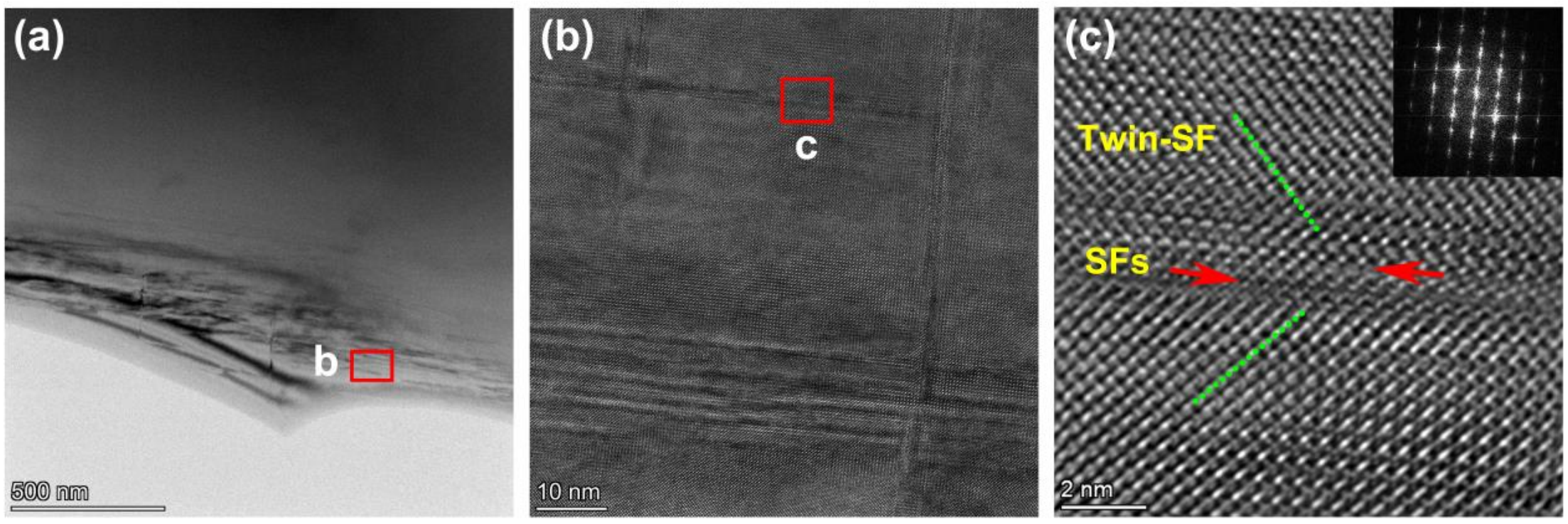

**Fig. 4** TEM images of $TiO_2$ sample with FP at 280 mA. (a,b) TEM images showing an area of the $TiO_2$ containing stacking faults. (c) HRTEM image showing a twin boundary decorated with stacking faults.

The phase evolution of the $TiO_2$ samples before and after FP was investigated. The XRD patterns shown in Fig. 5(a) revel that all samples retain a rutile crystal structure (PDF#73-2224), indicating that no phase transformation occurs in the sample after FP. Furthermore, electron paramagnetic resonance (EPR) spectroscopy was employed to study variations in oxygen vacancies ($O_V$). The signal intensity of $O_V$ (g = 2.003) is increased after FP (Fig. 5(b)), indicting an increase in oxygen vacancies after FP [30]. Notably, $TiO_2$ sample with FP at 280 mA contains the highest $O_V$ concentration among them. To further quantify the relative concentration of $O_V$, XPS analysis was performed (Fig. 5(c-e)). The content of $O_V$ is 57.88%, 65.31%, and 78.32% for pristine $TiO_2$ sample, and $TiO_2$ sample with FP at 230 and 280 mA, respectively (Fig. 5(f)). These results indicate that the concentration of oxygen vacancies in the $TiO_2$ sample is increased after FP, which also accounts for the color darkening of the sample after FP (inset of Fig. 1(b)) [31,32].

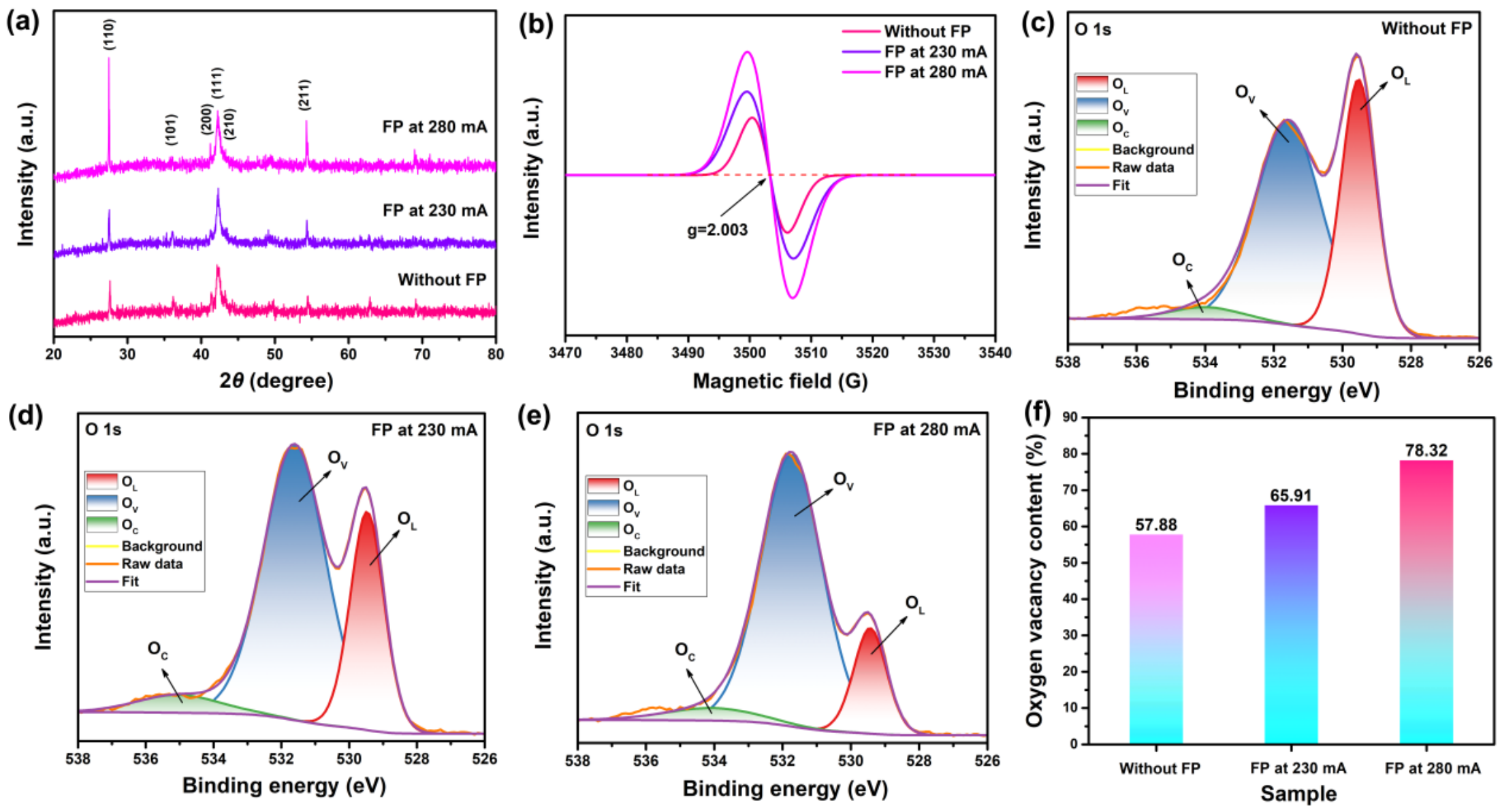

**Fig. 5** (a) XRD pattern. (b) EPR spectra. (c-e) XPS spectra. (f) Content of oxygen vacancies for different samples.

## 4. Discussions

The micropillar compression tests at room temperature show that pristine $TiO_2$ samples without FP experienced brittle failure, due to strong ionic bonds, high Peierls-Nabarro lattice resistance, and insufficient active slip systems at room temperature [33,34]. Microstructural characterization of the pristine $TiO_2$ confirms that, apart from point defects, there is virtually no line or planar defects in the sample (see supplementary Fig. S1). These defects are normally useful for providing the active nucleation sites for dislocations to facilitate the plastic deformation [16,35]. In contrast, samples with FP exhibited substantial improvement in their plasticity at room temperature, with compressive strain reaching 15.4% without catastrophic failure (Fig. 2(e-l)). Abundant shear bands formed and multiplied during compression to accommodate the plastic deformation.

It has been reported that the plasticity could be improved even at room temperature by the introduction of high-density dislocations into ceramic materials [14-16,35]. These high-density pre-existing dislocations circumvent the dislocation nucleation, which has been identified as one of the key bottlenecks for hindering dislocation plasticity in ceramics, particularly at room temperature [15,36]. Compared to the high stress level for dislocation nucleation, these pre-existing dislocations can glide at a much low shear stress level, preventing pre-mature fracture to happen while enabling the room-temperature dislocation mobility [6,34,37]. However, in our study, we found that FP did not generate a large number of dislocations. Therefore, the room-temperature plasticity observed in our work is unlikely to be predominantly driven by dislocation-assisted plastic deformation.

### 4.1 Introduction of defects after FP

Detailed microstructural characterization revealed that only a small number of stacking faults and twins, along with a large quantity of oxygen vacancies, were introduced into the $TiO_2$ sample after FP (Fig. 4 and 5). Moreover, virtually no dislocations were observed. The stacking faults and twins may be generated during the extraordinary non-equilibrium process in FP, resulting from the coupling of electric and thermal fields [38,39]. The oxygen vacancies primarily originate from the electrochemical reduction reaction [32,40,41].

### 4.2 Multiplication of defects during deformation

Detailed microstructural characterization of the $TiO_2$ pillars deformed to a strain of 8.0% and 14.7% revealed that more stacking faults and twin boundaries were produced under successive loading in the compression tests (Fig. 3). It is indicated that multiplication of defects (stacking faults and twins) occurs during deformation. We propose that oxygen vacancies play a very important role in the multiplication of such defects. Recently, Li et al. reported that oxygen vacancies could facilitate the nucleation and propagation of stacking faults and nanotwins during compression in their study on room-temperature plasticity of flash-sintered $TiO_2$ [24]. Okafor et al. also reported that oxygen vacancies could facilitate the nucleation of dislocations in the $SrTiO_3$ [42]. In our study, FP introduced a large concentration of oxygen vacancies, which is expected to lower the activation energy for the nucleation and propagation of stacking faults and twins, thereby promoting their continuous generation during compression. For the $TiO_2$ sample with FP at 280 mA, the higher concentration of oxygen vacancies results in lower activation energy for the nucleation and propagation of defects, thereby lowering the stress required to initiate plastic deformation (Fig. 2(i-l)). With increasing strain, however, the progressive accumulation and mutual interaction of these defects increasingly impede further defect-mediated deformation, resulting in the observed work hardening (Fig. 2(i)).

### 4.3 Mechanism of room-temperature plasticity

Stacking faults and twins are common planar defects in metals. It has been demonstrated that high-density stacking faults and twin boundaries enhance the strength of metal while retaining good ductility [43-46]. However, stacking faults and twins are rarely observed in most ceramics [47]. Moreover, it is extremely difficult to form the stacking faults and twins in ceramic materials [48], primarily due to the high energy penalty associated with severe distortion of strong covalent/ionic bonds. Recently, it was reported that fabrication of stacking faults and twins could significantly improve the plasticity of ceramics at room temperature [24,49-52]. Therefore, it is suggested that the room-temperature plasticity in our work is ascribed to the assistance of the stacking faults and twin boundaries. Figure 6 schematically illustrates the mechanism underlying room-temperature plastic deformation in $TiO_2$ with FP. Initially, a low density of stacking faults and twin boundaries, accompanied by abundant oxygen vacancies are introduced into the $TiO_2$ samples after FP. Upon room-temperature uniaxial compression, continuous generation and propagation of stacking faults and twin boundaries evolved into shear bands, facilitating the substantial room-temperature plastic deformability. Serrations observed on the stress-strain curves suggest the generation of stacking faults and twin boundaries during compression (Fig. 2(e-i)). The proliferation of stacking faults and twin boundaries is intimately linked to the high concentration of oxygen vacancies, which likely lowers the energy barrier for nucleation of these defects. Crucially, unlike conventional dislocation-dominated plasticity in ceramics, the room-temperature plasticity reported here operates via an oxygen-vacancy-mediated planar defect multiplication pathway.

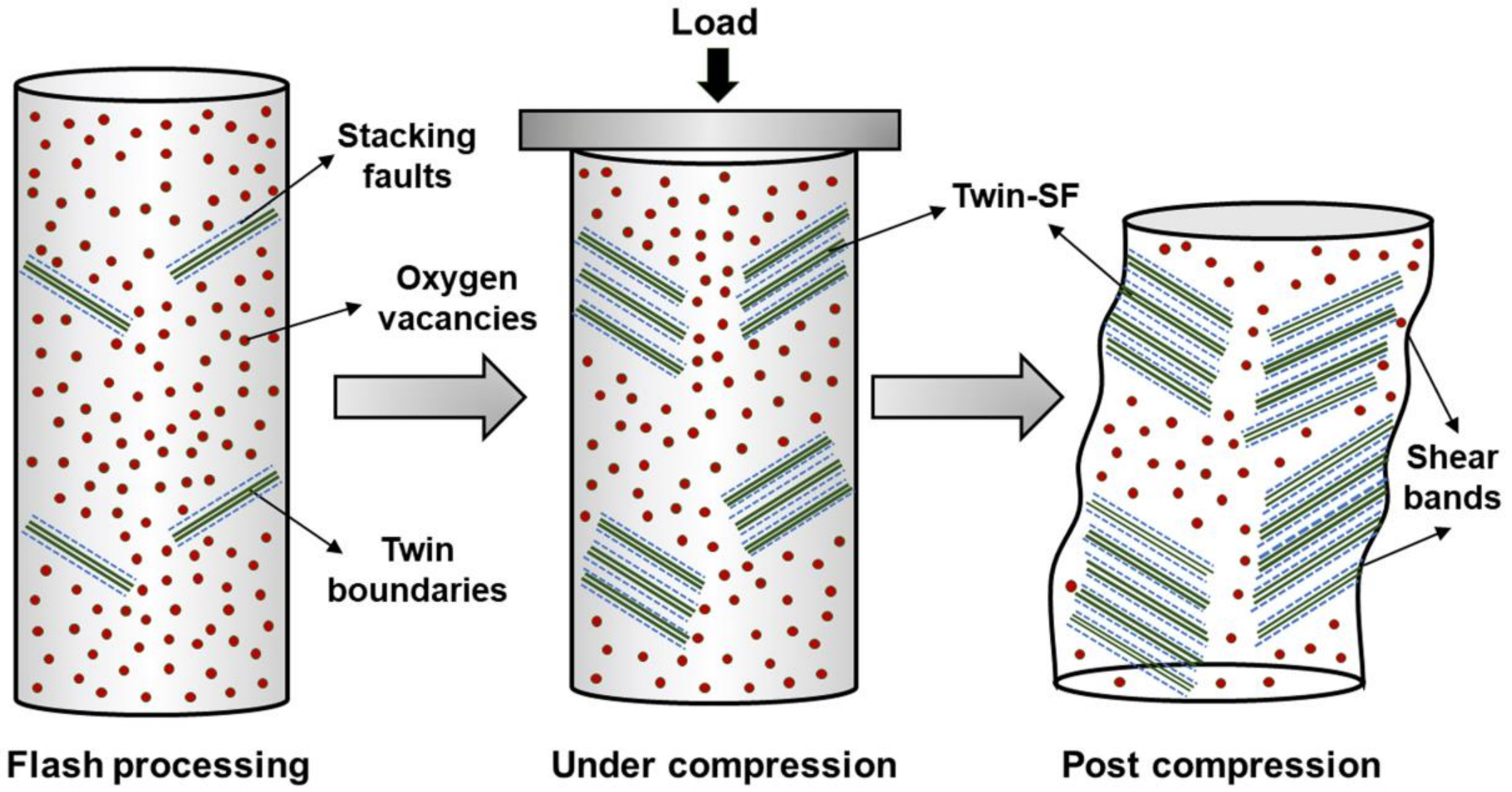

**Fig. 6** Schematic illustration of the mechanism underlying room-temperature plastic deformation in $TiO_2$ with FP.

## 5. Conclusions

This study presents a universal and practical strategy for achieving the room-temperature plasticity of ceramic materials through flash pressing (FP) based on the coupling of electric and thermal fields. The micropillar compression tests revealed a large plastic compressive strain of 15.4% at room temperature could be reached for the $TiO_2$ ceramics after FP. Differs from the dislocation-dominated plasticity in ceramics, the room-temperature plasticity in our study is ascribed to the multiplication of stacking faults and twin boundaries during the compression, which was facilitated by the high concentration of oxygen vacancies introduced by FP. As an emerging processing technique, FP holds great potential for tailoring the microstructure and optimizing the properties of ceramic materials.

**Acknowledgements**

This work was financially supported by the National Natural Science Foundation of China (grant No. 52472064), the Key Research and Development Projects of Shaanxi Province (grant No. 2025CY-YBXM-144), and the Fundamental Research Funds for the Central Universities, CHD (grant No. 300102315202).

**Author Contributions**

Xinghua Su: Writing-review & editing, Writing-original draft, Validation, Supervision, Methodology, Funding acquisition, Formal analysis, Conceptualization. Zhenyu Li: Visualization, Validation, Methodology, Investigation, Data curation. Xufei Fang: Writing-review & editing, Methodology. Wenjun Lu: Writing-review & editing. Mengyao Wang: Investigation. Peng Zhao: Project administration.

**Availability of data and materials**

The data that support the findings of this study are available from the corresponding author upon reasonable request.

**Competing interests**

The authors have no competing interests to declare that are relevant to the content of this article.